\documentclass[12pt]{iopartbis}
\usepackage{iopams}
\usepackage{graphicx}
\usepackage{color}

\begin{document}

\newcommand \be {\begin{equation}}
\newcommand \ee {\end{equation}}
\newcommand \bea {\begin{eqnarray}}
\newcommand \eea {\end{eqnarray}}
\newcommand \la {\langle}
\newcommand \ra {\rangle}
\newcommand \ve {\varepsilon}
\newcommand{\mC}{\mathcal{C}}
\newcommand{\mT}{\mathcal{T}}
\newcommand{\mE}{\mathcal{E}}
\newcommand{\mP}{\mathcal{P}}
\newcommand{\mR}{\mathcal{R}}
\newcommand{\mA}{\mathcal{A}}
\newcommand{\Lf}{\mathcal{L}}
\newcommand{\p}[1]{\left(#1\right)}

\title[]{Matrix Product representation of the stationary state of the open Zero Range Process}

\author{Eric Bertin}

\address{LIPHY, Univ.~Grenoble Alpes and CNRS, F-38000 Grenoble, France}

\author{Matthieu Vanicat}

\address{Faculty of Mathematics and Physics, University of Ljubljana,
Jadranska 19, SI-1000 Ljubljana, Slovenia}

\begin{abstract}
Many one-dimensional lattice particle models with open boundaries, like the paradigmatic Asymmetric Simple Exclusion Process (ASEP), have their stationary states represented in the form of a matrix product, with matrices that do not explicitly depend on the lattice site. In contrast, the stationary state of the open one-dimensional Zero-Range Process (ZRP) takes an inhomogeneous factorized form, with site-dependent probability weights. We show that in spite of the absence of correlations, the stationary state of the open ZRP can also be represented in a matrix product form, where the matrices are site-independent, non-commuting and determined from algebraic relations resulting from the master equation.
We recover the known distribution of the open ZRP in two different ways: first, using an explicit representation of the matrices and boundary vectors;
second, from the sole knowledge of the algebraic relations satisfied by these matrices and vectors.
Finally, an interpretation of the relation between the matrix product form and the inhomogeneous factorized form is proposed within the framework of hidden Markov chains.
\end{abstract}

\section{Introduction}

Computing analytically the stationary distribution of a non-equilibrium stochastic model is usually a very challenging task.
The general expression of the steady state using rooted trees expansion, see for instance \cite{Schnakenberg76}, generically involves a number of terms 
growing exponentially fast with the number of configurations of the system, and cannot be used in practice to compute efficiently 
physical quantities. It was nevertheless discovered that, in some very special cases, this apparent exponential complexity can be reduced to 
a polynomial computation. 
Indeed in the pioneering work \cite{DerridaEHP93}, the stationary distribution of the open Totally Asymmetric Simple Exclusion Process (TASEP) was
expressed in a matrix product form. This algebraic structure offered a very efficient framework to compute exactly 
the mean particle density and current, and to derive the phase diagram of the model.
It led to numerous developments and generalisations to other models with partially asymmetric hopping rate \cite{Sandow94},
with reaction-diffusion dynamics \cite{HinrichsenSP96,IsaevPR01,CrampeRRV16} 
or with several species of particles \cite{Karimipour99,Uchiyama08,EvansFM09,ProlhacEM09,AritaM13,CrampeMRV15,Vanicat17,FinnRV17}, 
see also the review articles \cite{Derrida07,BlytheE07} and references therein.
In fact it has been shown \cite{KrebsS97} that the stationary state of a large class of exclusion processes with open boundaries can be 
computed exactly using a homogeneous matrix ansatz.

While many one-dimensional models have a distribution taking a matrix product form, the Zero Range Process (ZRP) \cite{Spitzer70,EvansH05} with open boundaries stands alone with an inhomogeneous factorized distribution \cite{Schutz05}.
This result is a priori consistent with the fact that the ZRP has an unbounded number of configurations on each site, while the generic proof of existence of the matrix product state given in \cite{KrebsS97} assumes a finite number of local configurations.
However, matrix product distributions have recently been found for different types of generalized ZRP with periodic boundary conditions \cite{ChatterjeeM17,KunibaO17,KunibaM17}.
These results thus raise the question whether the open ZRP may also fall into the class of models that can be described by a matrix product state.

In this short note, we show that the inhomogeneous factorized distribution 
of the open ZRP can also be obtained from the matrix-product ansatz, without prior knowledge of the factorization property. 
The distribution can be determined either from explicit representations of the matrices and boundary vectors, or using only the 
algebraic relations satisfied by these matrices and vectors.
We also discuss the connection of these results with the recently introduced Hidden Markov Chain formalism for matrix-product distributions
\cite{AngelettiTSP13,AngelettiEPL13,AngelettiJSP14}.

\section{Open Zero Range Process}

\subsection{Definition of the model}

The ZRP \cite{Spitzer70,EvansH05} is one of the simplest interacting particle lattice models, in which the probability to move a particle from one site to another only depends on the number of particles on the departure site.
Due to its simplicity, the ZRP has a factorized steady-state probability distribution \cite{EvansH05}, and this factorization property is preserved (at least in one-dimension) when considering open boundaries connected to particle reservoirs \cite{Schutz05}, or when studying large deviations of the current
\cite{HarrisRS05}.

The open one-dimensional ZRP is defined as follows \cite{Schutz05}.
An arbitrary number $n_i$ of particles can lie on any of the sites
$i=1,\dots,L$ of the lattice.
The probability per unit time to transfer a particle from site $i$ to site $i+1$ (resp. $i-1$) is $q u(n_i)$ [resp. $(1-q) u(n_i)$], where $0 < q < 1$ is a parameter of the model (for later computational convenience, we exclude the limit cases $q=0$ and $q=1$). The function $u(n)$ is the probability per unit time that a particle is moved to a neighboring site, given that there is $n>0$ particles on the departure site. For convenience, we also set $u(0)=0$.

In addition, boundary sites $i=1$ and $i=L$ exchange particles with reservoirs.
The `left' reservoir injects a particle on site $i=1$ with a rate $\alpha$. A particle situated on site $i=1$ is transfered to the reservoir with a rate $(1-q)u(n_1)$.
Symmetrically, the `right' reservoir injects a particle on site $i=L$ with a rate $\beta$, and withdraws a particle from this site with a rate $q u(n_L)$.
Note that slightly more general transition rates with the reservoirs have been considered in \cite{Schutz05}. Here, for the sake of simplicity, we choose the same bias $q$ and $(1-q)$ as in the bulk for the interaction with the reservoirs.

The steady-state distribution $P(n_1,\dots,n_L)$ of the open ZRP is known 
\cite{Schutz05}, and given by the inhomogeneous factorized form
\be \label{eq:inhomog:factor:P}
P(n_1,\dots,n_L) = \prod_{k=1}^L p_k(n_k)
\ee
with
\be \label{eq:marginal:pk}
p_k(n) = \frac{z_k^n}{\mathcal{Z}_k} \, \prod_{m=1}^n \frac{1}{u(m)}
\ee
where $\mathcal{Z}_k$ is a normalization factor ensuring $\sum_{n=0}^{\infty} p_k(n)=1$. By convention, the product in Eq.~(\ref{eq:marginal:pk}) is equal to $1$ when $n=0$. The local `fugacity' $z_k$ is given by \cite{Schutz05}
\be \label{eq:zk}
z_k = \frac{\alpha}{q} \left(\frac{q}{1-q}\right)^k+\left(1- \left(\frac{q}{1-q}\right)^k\right)\frac{\alpha q^L-\beta(1-q)^L}{q^{L+1}-(1-q)^{L+1}} \,.
\ee
In the following, we investigate whether the open ZRP can alternatively be solved using the standard matrix product ansatz method.
We thus start by writing down explicitly the master equation of the model.

\subsection{Stationary master equation}

The master equation governing the probability distribution $P(\{n_j\},t)$ of the open ZRP can be written formally as
\be \label{eq:MEq:ZRP}
\hspace{-2cm}\frac{\partial P}{\partial t}(\{n_j\},t)
= \sum_{n_1',\dots,n_L'} \Big[ \mathcal{W}(\{n_j\}|\{n_j'\}) P(\{n_j'\},t)
- \mathcal{W}(\{n_j'\}|\{n_j\}) P(\{n_j\},t) \Big] \,,
\ee
where the transition rates can be decomposed into a sum of local operators,
\be
\hspace{-2cm}\mathcal{W}(\{n_j'\}|\{n_j\}) = \sum_{i=1}^{L-1} \mathcal{M}_{i,i+1}(\{n_j'\}|\{n_j\})
+ \mathcal{B}_1(\{n_j'\}|\{n_j\}) + \mathcal{B}_L(\{n_j'\}|\{n_j\}) \,.
\ee
The bulk rate $\mathcal{M}_{i,i+1}(\{n_j'\}|\{n_j\})$ is given by
\begin{align} \nonumber
\mathcal{M}_{i,i+1}(\{n_j'\}|\{n_j\}) &= \left[ \prod_{j \ne i,i+1} \delta_{n_j',n_j} \right]
\left[ q u(n_i) \,\delta_{n_i-1,n_i'} \,\delta_{n_{i+1}+1,n_{i+1}'} \right. \\
 & \qquad \qquad \qquad \qquad \left. + (1-q) u(n_{i+1}) \,\delta_{n_{i+1}-1,n_{i+1}'} \,\delta_{n_i+1,n_i'} \right]
\end{align}
while the boundary operators are defined as
\begin{align} \nonumber
\mathcal{B}_1(\{n_j'\}|\{n_j\}) &= \left[ \prod_{j=2}^L \delta_{n_j',n_j} \right]
\left[ (1-q)\, u(n_1) \, \delta_{n_1-1,n_1'}
+ \alpha \, \delta_{n_1+1,n_1'} \right]\\
\mathcal{B}_L(\{n_j'\}|\{n_j\}) &= \left[ \prod_{j=1}^{L-1} \delta_{n_j',n_j} \right]
\left[ q\, u(n_L) \, \delta_{n_L-1,n_L'} + \beta \, \delta_{n_L+1,n_L'} \right]
\end{align}
where $\delta_{n,n'}$ is the Kronecker delta symbol.
This form of the master equation is useful to explore the matrix product ansatz solution of the steady-state distribution, as explained in the next section.

\section{Matrix product ansatz solution}

\subsection{Reformulation of the master equation using matrix product ansatz}

In the following, we look for a stationary solution of the master equation (\ref{eq:MEq:ZRP}) in the Matrix Product Ansatz form:
\be \label{eq:MPA}
P_{\rm st}(n_1,\dots,n_L) = \frac{1}{Z} \, \langle W| R(n_1) \dots R(n_L) |V\rangle
\ee
where $R(n)$ is a matrix-valued function of the integer variable $n$,
and $\langle W|$ and $|V\rangle$ are boundary vectors;
$Z$ is the normalization constant $Z=\langle W|\mE^L|V\rangle$,
where the matrix $\mE$ is defined as
\be \label{eq:def:E}
\mE = \sum_{n=0}^{\infty} R(n) \,.
\ee
Considering again the stationary master equation, the `telescopic' relation (see, e.g., \cite{BlytheE07}) involving the bulk transition rate $\mathcal{M}_{i,i+1}$ takes the form 
\begin{align} \label{eq:telescopic}
& q u(n_1+1) R(n_1+1) R(n_2-1) + (1-q) u(n_2+1) R(n_1-1) R(n_2+1) \\
& - q u(n_1) R(n_1) R(n_2) -(1-q) u(n_2) R(n_1) R(n_2)
= R(n_1) \overline{R}(n_2) - \overline{R}(n_1) R(n_2)
\nonumber
\end{align}
where $\overline{R}(n)$ is another matrix-valued function to be determined.
Note that $\overline{R}(n)$ does not explicitly appear in the distribution 
$P_{\rm st}(n_1,\dots,n_L)$, and cancels out when summing over the sites in
the stationary version of the master equation (\ref{eq:MEq:ZRP}).

In addition, relations coming from the boundary conditions involving the transition rates $\mathcal{B}_1$ and $\mathcal{B}_L$ respectively read
\begin{align} \nonumber
& (1-q) u(n_1+1)\, \langle W| R(n_1+1)
+ \alpha \, \langle W| R(n_1-1)\\
& \qquad \qquad \qquad - (1-q) u(n_1) \, \langle W| R(n_1)
- \alpha \, \langle W| R(n_1)
= \langle W| \overline{R}(n_1)
\label{eq:algeb:left}
\end{align}
and
\begin{align} \nonumber
& q u(n_L+1) R(n_L+1) |V\rangle + \beta R(n_L-1) |V\rangle \\
&  \qquad \qquad \qquad - q u(n_L) R(n_L)|V\rangle
- \beta R(n_L)|V\rangle = - \overline{R}(n_L)|V\rangle \,.
\label{eq:algeb:right}
\end{align}
It is convenient to reformulate Eqs.~(\ref{eq:telescopic}),
(\ref{eq:algeb:left}) and (\ref{eq:algeb:right}) by introducing the
following change of function:
\be \label{eq:RfK}
R(n) = f(n) K(n), \qquad \overline{R}(n) = f(n) \overline{K}(n)
\ee
where $K(n)$ and $\overline{K}(n)$ are matrix-valued functions to be determined, and
\be
f(n) = \prod_{m=1}^n \frac{1}{u(m)} \;.
\ee
Eqs.~(\ref{eq:telescopic}), (\ref{eq:algeb:left}) and (\ref{eq:algeb:right}) then simplify to
\begin{align}
\label{eq:telescopic:ZRP}
& u(n_1) \, \Big[ (1-q) K(n_1-1) K(n_2+1) - q K(n_1) K(n_2) \Big] \\
\nonumber
& + u(n_2) \, \Big[ q K(n_1+1) K(n_2-1) - (1-q) K(n_1) K(n_2) \Big]
= K(n_1) \overline{K}(n_2) - \overline{K}(n_1) K(n_2) \,, \\
& \nonumber \, \\
\label{eq:algeb:left:ZRP}
& \Big[ (1-q) \, \langle W| K(n_1+1) - \alpha \, \langle W | K(n_1) \Big] \\
\nonumber
& \qquad \qquad + u(n_1) \, \Big[ \alpha \, \langle W| K(n_1-1) -(1-q) \, \langle W| K(n_1)
\Big] = \langle W | \overline{K}(n_1) \,, \\
& \nonumber \, \\
\label{eq:algeb:right:ZRP}
& \Big[q K(n_L+1) |V\rangle - \beta K(n_L) |V\rangle \Big] \\
\nonumber
& \qquad \qquad + u(n_L) \, \Big[ \beta K(n_L-1) |V\rangle - q K(n_L) |V\rangle \Big]
= -\overline{K}(n_L) |V\rangle \,.
\end{align}
In the following, we look for a parameterization of the matrix-valued function
$K(n)$ allowing for a simpler reformulation of
Eqs.~(\ref{eq:telescopic:ZRP}), (\ref{eq:algeb:left:ZRP}) and (\ref{eq:algeb:right:ZRP}).

\subsection{Parameterization of the matrix $K(n)$ and algebraic relations}

The appearance of the terms $K(n_1-1) K(n_2+1)$, $K(n_1) K(n_2)$
and $K(n_1+1) K(n_2-1)$ in Eq.~(\ref{eq:telescopic:ZRP}) suggests that
$K(n)$ could have an exponential dependence on $n$, say $K(n)=A^n$,
where $A$ is a matrix. Such a simple form converts the terms $K(n_1-1) K(n_2+1)$ and $K(n_1+1) K(n_2-1)$ into $K(n_1) K(n_2)$, thus greatly simplifying Eq.~(\ref{eq:telescopic:ZRP}). However, such a pure exponential form leads to commuting matrices $K(n)$ and $K(n')$, and is not consistent with a nonuniform density profile.
We thus choose a slightly more involved parameterization the matrix $K(n)$ of the form
\be \label{eq:param:Kn}
K(n) = B A^n
\ee
with two unknown matrices $A$ and $B$.
A careful inspection of the bulk equation (\ref{eq:telescopic:ZRP}) shows that it is satisfied if the matrices $A$ and $B$ obey a commutation relation of the form
\be \label{eq:commut:ZRP}
qAB-(1-q)BA=cB \,,
\ee
with $c$ an unknown real parameter, provided $\overline{K}(n)$ is chosen as
\be
\overline{K}(n) = c\, u(n) B A^{n-1} + c' B A^n \,,
\ee
where $c'$ is also an arbitrary real parameter.
Using these relations in the boundary equations (\ref{eq:algeb:left:ZRP})
and (\ref{eq:algeb:right:ZRP}), we end up with the four equations
\bea
q\langle W| A &=& (c+c'+\alpha) \langle W| \,,\\
q\langle W| A &=& \alpha \langle W| \,,\\
q A |V\rangle &=& (\beta-c') |V\rangle \,,\\
q A |V\rangle &=& (\beta+c) |V\rangle \,.
\eea
Consistency of the above four equations implies $c'=-c$.

\subsection{Explicit representation of the matrices $A$ and $B$}

We would like to construct an explicit representation of the matrices $A$ and $B$ and of the boundary vectors $\langle W|$ and $|V\rangle$ satisfying
the relations
\be
\label{eq:algeb}
\hspace{-2cm} qAB-(1-q)BA=cB \,, \qquad q\langle W|A = \alpha \langle W| \,,
\qquad qA |V\rangle = (\beta+c)|V\rangle.  
\ee
This construction may require to select a specific value of the parameter $c$,
which is up to now arbitrary.
For reasons that will become clear below, we consider finite-dimensional representations of dimension $L+1$, where $L$ is the number of sites of the lattice.
The vector space is spanned by the $L+1$ vectors $\{|k\rangle\}_{k=0}^L$.
We choose the following parameterizations for the matrices $A$, $B$, and the vectors $\langle W|$ and $|V\rangle$
\begin{align}
\label{eq:param:B:ZRP}
B &= \sum_{k=1}^L |k-1\rangle \langle k| = \begin{pmatrix}
                                            0 & 1 & 0 & \cdots & 0 \\
                                            \vdots & \ddots & \ddots & \ddots & \vdots \\
                                            \vdots & & \ddots & \ddots & 0 \\
                                            \vdots & & & \ddots & 1 \\
                                            0 & \cdots & \cdots & \cdots & 0
                                           \end{pmatrix} \,, \\
\nonumber \, \\
\label{eq:param:A:ZRP}
A &= \sum_{k=0}^L z_k |k\rangle \langle k| = \begin{pmatrix}
                                              z_0 & 0 & \cdots & 0 \\
                                              0 & z_1 & \ddots & \vdots \\
                                              \vdots & \ddots & \ddots & 0 \\
                                              0 & \cdots & 0 & z_L
                                             \end{pmatrix} \,,\\
\langle W| &= \langle 0| = \begin{pmatrix}
                            1 & 0 & \cdots & 0 
                           \end{pmatrix}, \qquad
|V\rangle = |L\rangle = \begin{pmatrix}
              0 \\ \vdots \\ 0 \\ 1
             \end{pmatrix}.
\end{align}
The sequence $(z_0,\dots,z_L)$ has to satisfy 
\begin{equation} \label{eq:recur}
 q z_k -(1-q)z_{k+1} = c, \quad \mbox{for} \quad k=0,\dots,L-1
\end{equation}
with the boundary conditions
\begin{equation} \label{eq:boundary:zk}
 q z_0 = \alpha, \qquad q z_L = \beta + c.
\end{equation}
The solution of Eq.~(\ref{eq:recur}) satisfies the boundary condition Eq.~(\ref{eq:boundary:zk}) only if (assuming $q \ne \frac{1}{2}$)
\begin{equation} \label{eq:value:c}
 c = (2q-1) \frac{\alpha q^L-\beta(1-q)^L}{q^{L+1}-(1-q)^{L+1}} \,.
\end{equation}
The case $q=\frac{1}{2}$ is obtained by taking the limit of the above expression when $q \to \frac{1}{2}$.
Then the solution of Eqs.~(\ref{eq:recur}) and (\ref{eq:boundary:zk}) is precisely Eq.~(\ref{eq:zk}).
%
We are now able to compute the stationary probability distribution
$P(n_1,\dots,n_L)$.
Given that
\be
BA^n = \sum_{k=1}^L z_k^n |k-1\rangle \langle k|
\ee
one finds
\begin{align} \nonumber
\langle W|BA^{n_1}BA^{n_2}\dots BA^{n_L}|V\rangle
&= \sum_{k_1,\dots,k_L=1}^L z_{k_1}^{n_1} \dots z_{k_L}^{n_L} \,
\langle W|k_1-1\rangle \langle k_1|k_2-1\rangle \dots \\
\nonumber
& \qquad \qquad \qquad \qquad \qquad \qquad\dots \langle k_{L-1}|k_L-1\rangle
\langle k_L|V\rangle \\
\nonumber
&= \sum_{k_1,\dots,k_L=1}^L z_{k_1}^{n_1} \dots z_{k_L}^{n_L} \, \delta_{0,k_1-1} \,
\delta_{k_1,k_2-1} \dots \delta_{k_{L-1},k_L-1} \, \delta_{k_L,L} \\
&= z_1^{n_1}z_2^{n_2}\dots z_L^{n_L} \,.
\label{eq:WBAV}
\end{align}
Since from Eqs.~(\ref{eq:RfK}) and (\ref{eq:param:Kn}), $R(n)=f(n) B A^n$,
we obtain using Eq.~(\ref{eq:WBAV}) that the distribution $P(n_1,\dots,n_L)$
given by the matrix product form Eq.~(\ref{eq:MPA}) boils down
to the inhomogeneous factorized form given in
Eqs.~(\ref{eq:inhomog:factor:P}) and (\ref{eq:marginal:pk}), previously derived in \cite{Schutz05}.

\subsection{Alternative derivation of the distribution from the sole algebraic relations}

The interest of having an explicit representation of the matrices $A$ and $B$
and vectors $\langle W|$ and $|V\rangle$ is twofold: first, one is then sure 
that matrices and vectors satisfying the algebraic relations given in
(\ref{eq:algeb})
do exist; second, having an explicit form is obviously convenient to perform calculations in practice.
It is known, however, that in the framework of the ASEP model it is possible to determine the 
probability distribution using only the algebraic relations satisfied by the matrices and vectors,
without having an explicit representation at hand \cite{Derrida07,CrampeMRV15,CrampeRRV16,Vanicat17}.
It is interesting to see whether such an algebraic approach also works for the open ZRP.

To evaluate the probability distribution $P(n_1,\dots,n_L)$, we need to compute quantities of the form $\langle W|BA^{n_1}BA^{n_2}\dots BA^{n_L}|V\rangle$.
This was done in Eq.~(\ref{eq:WBAV}) using an explicit representation of the matrices $A$, $B$ and vectors $\langle W|$ and $|V\rangle$.
Now we would like to evaluate the quantity $\langle W|BA^{n_1}BA^{n_2}\dots BA^{n_L}|V\rangle$ using only the algebraic relations (\ref{eq:algeb}).
In particular, we need to use the fact that, according to Eq.~(\ref{eq:algeb}), $\langle W|$ and $|V\rangle$ are eigenvectors of the matrix $A$.
To use this property, we have to transform the product $BA^{n_1}BA^{n_2}\dots BA^{n_L}$ by `pulling' all the matrices $A$ either to the right or to the left, using the relation $qAB-(1-q)BA=cB$.
We start by rewriting the latter equation as
\be
AB = \lambda BA + \mu B \,, \qquad \lambda \equiv \frac{1-q}{q} \,,
\quad \mu \equiv \frac{c}{q} \,.
\ee
It is easy to show, by recursion, that for all integer $n \ge 0$,
\be \label{eq:expansion:AnB}
A^n B = \sum_{k=0}^n \binom{n}{k} \, \lambda^k \mu^{n-k} B A^k \,.
\ee
We now reexpress the product $BA^{n_1}BA^{n_2}\dots BA^{n_L}$ as a linear combination of products of the form $B^L A^k$, with $0\le k \le \sum_{i=1}^L n_i$.
This can be done by repeatedly applying Eq.~(\ref{eq:expansion:AnB})
in the product $BA^{n_1}BA^{n_2}\dots BA^{n_L}$, which leads to
\begin{align} \nonumber
BA^{n_1}BA^{n_2}\dots BA^{n_L}
&= \sum_{k_1=0}^{n_1} \sum_{k_2=0}^{k_1+n_2} \sum_{k_3=0}^{k_2+n_3} \dots
\sum_{k_{L-1}=0}^{k_{L-2}+n_{L-1}}
\binom{n_1}{k_1} \lambda^{k_1} \mu^{n_1-k_1} \times\\ \nonumber
& \times \binom{k_1+n_2}{k_2} \lambda^{k_2} \mu^{k_1+n_2-k_2}
\binom{k_2+n_3}{k_3} \lambda^{k_3} \mu^{k_2+n_3-k_3} \times \dots\\
& \dots \times \binom{k_{L-2}+n_{L-1}}{k_{L-1}}
\lambda^{k_{L-1}} \mu^{k_{L-2}+n_{L-1}-k_{L-1}}\, B^L A^{k_{L-1}+n_L} \,.
\label{eq:expansion:BAB}
\end{align}
Using the fact that $|V\rangle$ is an eigenvector of $A$, see Eq.~(\ref{eq:algeb}), we have
\begin{align} \nonumber
&\langle W|BA^{n_1}BA^{n_2}\dots BA^{n_L}|V\rangle
= \langle W|B^L|V\rangle \left(\frac{\beta+c}{q}\right)^{n_L} \times\\ \nonumber
&\qquad\qquad\qquad \times \sum_{k_1=0}^{n_1} \sum_{k_2=0}^{k_1+n_2} \sum_{k_3=0}^{k_2+n_3} \dots
\sum_{k_{L-1}=0}^{k_{L-2}+n_{L-1}}
\binom{n_1}{k_1} \lambda^{k_1} \mu^{n_1-k_1} \times\\ \nonumber
&\qquad\qquad\qquad \times \binom{k_1+n_2}{k_2} \lambda^{k_2} \mu^{k_1+n_2-k_2}
\binom{k_2+n_3}{k_3} \lambda^{k_3} \mu^{k_2+n_3-k_3} \times \dots\\
&\qquad\qquad\qquad \dots \times \binom{k_{L-2}+n_{L-1}}{k_{L-1}} \lambda^{k_{L-1}} \mu^{k_{L-2}+n_{L-1}-k_{L-1}}\; \left(\frac{\beta+c}{q}\right)^{k_{L-1}} \,.
\label{eq:expansion:WBABV}
\end{align}
A careful look at Eq.~(\ref{eq:expansion:WBABV}) then shows that it can be factorized as
\be \label{eq:factorized:right}
\hspace{-2cm}\langle W|BA^{n_1}BA^{n_2}\dots BA^{n_L}|V\rangle
= \langle W|B^L|V\rangle \prod_{k=1}^L \left[ \lambda^{L-k}
\left(\frac{\beta+c}{q}\right) + \mu \sum_{j=0}^{L-k-1} \lambda^j \right]^{n_k}
\ee
with the convention that the empty sum is equal to zero.
A similar procedure can also be applied to use the fact that the vector
$\langle W|$ is an eigenvector of $A$.
In this case, we need to `push' all the $A$'s to the left.
Writing
\be
BA = \tilde{\lambda} AB + \tilde{\mu} B \,, \qquad
\tilde{\lambda} \equiv \frac{1}{\lambda} \,, \quad
\tilde{\mu} \equiv -\frac{\mu}{\lambda} \,,
\ee
we end up with
\be \label{eq:expansion:BAn}
B A^n = \sum_{k=0}^n \binom{n}{k}\, \tilde{\lambda}^k \tilde{\mu}^{n-k} A^k B \,.
\ee
Expanding $\langle W|BA^{n_1}BA^{n_2}\dots BA^{n_L}|V\rangle$ using
Eq.~(\ref{eq:expansion:BAn}) as well as $\langle W|A = (\alpha/q)\langle W|$,
see Eq.~(\ref{eq:algeb}), the resulting expansion can be factorized into
\be \label{eq:factorized:left}
\langle W|BA^{n_1}BA^{n_2}\dots BA^{n_L}|V\rangle
= \langle W|B^L|V\rangle \prod_{k=1}^L \left[ \tilde{\lambda}^k
\, \frac{\alpha}{q} + \tilde{\mu} \sum_{j=0}^{k-1} \tilde{\lambda}^j \right]^{n_k}.
\ee
Up to now, the value of the parameter $c$ has been left unspecified,
but we have to ensure that the two expressions obtained for
$\langle W|BA^{n_1}BA^{n_2}\dots BA^{n_L}|V\rangle$ are identical.
Identifying Eqs.~(\ref{eq:factorized:right}) and (\ref{eq:factorized:left}),
we obtain for all $k=1,\dots,L$,
\be \label{eq:left:right:consistent}
\lambda^{L-k}
\left(\frac{\beta+c}{q}\right) + \mu \sum_{j=0}^{L-k-1} \lambda^j
= \tilde{\lambda}^k \, \frac{\alpha}{q} + \tilde{\mu} \sum_{j=0}^{k-1} \tilde{\lambda}^j \,.
\ee
Using $\tilde{\lambda} = 1/\lambda$, $\tilde{\mu} = -\mu/\lambda$
as well as $\lambda=(1-q)/q$ and $\mu=c/q$, one finds after
summing the geometric series that the $k$-dependence cancels out,
and one recovers the value of $c$ given in Eq.~(\ref{eq:value:c}).
Then both sides of Eq.~(\ref{eq:left:right:consistent}) identify with
the factor $z_k$ given in Eq.~(\ref{eq:zk}).
The unknown factor $\langle W|B^L|V\rangle$ in Eqs.~(\ref{eq:factorized:right}) and (\ref{eq:factorized:left}) cancels out when normalizing the distribution
$P(n_1,\dots,n_L)$, and one recovers the factorized expression given in
Eqs.~(\ref{eq:inhomog:factor:P}) and (\ref{eq:marginal:pk}).

\section{Interpretation in terms of Hidden Markov chain}

The relation between the matrix product form and the inhomogeneous factorized form can actually be put in a broader perspective using the recently introduced Hidden Markov Chain representation of distributions having a matrix product form
\cite{AngelettiTSP13,AngelettiEPL13,AngelettiJSP14}
---see also, e.g., \cite{Cappe05} for a general introduction on the Hidden Markov Chain formalism in the context of theoretical signal processing.
This framework assumes that for all $(i,j)$ and all $n$, $R_{ij}(n) \ge 0$. This property is satisfied by the representation $R(n)=f(n)BA^n$ with matrices $B$ and $A$ defined by Eqs.~(\ref{eq:param:B:ZRP}), (\ref{eq:param:A:ZRP})
and (\ref{eq:zk}).
It is useful to introduce the matrix of distributions $\mP(x)$ through the relation
\begin{equation}
R_{ij}(n) = \mE_{ij} \mP_{ij}(n),
\end{equation}
[where the matrix $\mE$ is defined in Eq.~(\ref{eq:def:E})] so that $\mP_{ij}(n)$ can be interpreted for fixed $i,j$ as a probability distribution of the variable $n$,
normalized to $1$. 
The distribution $\mP_{ij}(n)$ is uniquely defined for all $(i,j)$ such that $\mE_{ij} \ne 0$. If $\mE_{ij} = 0$, $\mP_{ij}(n)$ is arbitrary and plays no role.
To interpret the joint probability Eq.~(\ref{eq:MPA}) in the framework of Hidden Markov Chains, we introduce a Markov chain $\Gamma\in \{0,\dots L\}^{L+1}$ such that \cite{AngelettiTSP13,AngelettiEPL13}
\begin{align}
\label{eqn:trans0}
{\rm Pr}(\Gamma_1 = i,\Gamma_{L+1} = f) &= \frac{W_i \, (\mE^L)_{if} V_f}{\langle W|\mE^L | V \rangle } \; , \\
 \label{eq:transition}
{\rm Pr}(\Gamma_{k+1}= j | \Gamma_{k}=i, \, \Gamma_{L+1} = f ) &= \mE_{ij} \frac{(\mE^{L-k})_{jf}}{(\mE^{L-k+1})_{if}} \; .
\end{align}
The Markov chain $\Gamma$ is non-homogeneous and of a nonstandard type, due to the dependence on the final state $\Gamma_{L+1}$, which enters the transition rate ${\rm Pr}(\Gamma_{k+1}=j|\Gamma_{k}=i,\, \Gamma_{L+1}=f)$. Note that for $k=L$,
${\rm Pr}(\Gamma_{k+1}=j|\Gamma_{k}=i,\, \Gamma_{L+1}=f)=1$ if $j=f$ and $0$ otherwise.
Initial and final states of the Markov chain are randomly drawn from the probability distribution ${\rm Pr}(\Gamma_1=i,\Gamma_{L+1}=f)$.
From Eqs.~(\ref{eqn:trans0}) and (\ref{eq:transition}),
the probability $\kappa(\Gamma)$ of a chain $\Gamma=(\Gamma_1,\dots,\Gamma_{L+1})$ is obtained as
\begin{equation} \label{eq:kappa}
\kappa(\Gamma) = \frac{W_{\Gamma_1} V_{\Gamma_{L+1}}}{\langle W| \mE^L |V\rangle}\,
\mE_{\Gamma_1 \Gamma_2}\, \mE_{\Gamma_2 \Gamma_3}  \dots \mE_{\Gamma_L \Gamma_{L+1}} \,.
\end{equation}
For a given $\Gamma$, the random variables $(n_1,\ldots,n_L)$ are independent but non-identically distributed, 
with a probability distribution depending on $\Gamma$:
\begin{equation} \label{eq:X:gamma}
P(n_1,\dots,n_L|\Gamma) = \prod_{k=1}^L \mP_{\Gamma_k \Gamma_{k+1}}(n_k) \,,
\end{equation}
and the full distribution $P(n_1,\dots,n_L)$ reads
\be \label{eq:Psum:gamma}
P(n_1,\dots,n_L) = \sum_{\Gamma} \kappa(\Gamma) \, P(n_1,\dots,n_L|\Gamma) \,.
\ee
In the case of the open ZRP,
\be
\mE = \sum_{k=1}^L \frac{1}{1-z_k} \, |k-1\rangle \langle k|
\ee
so that
\be
\mE^L = \left( \prod_{k=1}^L \frac{1}{1-z_k} \right) \, |0\rangle \langle L| \,.
\ee
From Eqs.~(\ref{eqn:trans0}), (\ref{eq:transition}) and (\ref{eq:kappa}),
the only chain $\Gamma$ for which the probability
$\kappa(\Gamma)$ is nonzero is given by $\Gamma_k=k-1$, for $k=1,\dots,L+1$.
It follows that the sum in Eq.~(\ref{eq:Psum:gamma}) reduces to a single term, so that the distribution $P(n_1,\dots,n_L)$ is factorized.
In contrast, a similar treatment of the ASEP model for instance would yield a sum over a large number of different chains $\Gamma$. Although apparently complicated, such a representation of the distribution would be useful for instance to investigate the fluctuations of the total number of particles, where the ergodic (or non-ergodic) properties of the Hidden Markov chain $\Gamma$ play a key role
\cite{AngelettiEPL13,AngelettiJSP14}.

\section{Conclusion}

The derivation presented here provides an alternative way to derive the steady-state distribution of the open ZRP. 
The path followed is not necessarily easier than the one originally followed in \cite{Schutz05}, but rather provides a different perspective on the derivation.
Our aim was to show that the matrix-product ansatz approach
may also be valid to determine the steady-state distribution of boundary-driven one-dimensional lattice models with unbounded number of particles on each site
(see also, e.g., \cite{ChatterjeeM17,KunibaO17,KunibaM17} for other applications of 
the matrix ansatz to models with unbounded local number of particles and periodic boundary conditions).
It also shows that in some specific cases, the homogeneous matrix product form and the inhomogeneous factorized form 
(i.e., with local distributions that explicitly depend on the site), may be two sides of the same coin.
More generally, the matrix-product form can be reformulated as a mixture of inhomogeneous factorized distributions, 
within the hidden Markov chain framework \cite{AngelettiTSP13,AngelettiEPL13,AngelettiJSP14}.
As for future work, the present matrix product ansatz approach might be useful to investigate the stationary distribution of more general 
models like an open version of the continuous mass transport model introduced in \cite{Evans04a,Evans04b,Majumdar05}.

\section*{Acknowledgments}

The authors are grateful to V. Lecomte and E. Ragoucy for interesting discussions. 
M.V. acknowledges financial support from ERC grant No. 694544. 
Both authors also acknowledge financial support from the grant IDEX-IRS `PHEMIN' of the Universit\'{e} Grenoble Alpes.

\bigskip



\begin{thebibliography}{99}

\bibitem{Schnakenberg76}
J. Schnakenberg, Rev. Mod. Phys. {\bf 48} (1976) 571.

\bibitem{DerridaEHP93}
B. Derrida, M.R. Evans, V. Hakim, V. Pasquier, J. Phys. A: Math. Theor. {\bf 26} (1993) 1493.

\bibitem{Sandow94}
S. Sandow, Phys. Rev. E {\bf 50} (1994) 2660.

\bibitem{HinrichsenSP96}
H. Hinrichsen, S. Sandow, I. Peschel, J. Phys. A: Math. Gen. {\bf 29} (1996) 2643.

\bibitem{IsaevPR01}
A.P. Isaev, P.N. Pyatov, V. Rittenberg, J. Phys. A: Math. Gen. {\bf 34} (2001) 5815.

\bibitem{CrampeRRV16}
N. Crampe, E. Ragoucy, V. Rittenberg, M. Vanicat, Phys. Rev. E {\bf 94} (2016) 032102.

\bibitem{Karimipour99}
V. Karimipour, Phys. Rev. E {\bf 59} (1999) 205.

\bibitem{Uchiyama08}
M. Uchiyama, Chaos, Solitons and Fractals {\bfseries 35} (2008) 398.

\bibitem{EvansFM09}
M.R. Evans, P.A. Ferrari, K. Mallick, J. Stat. Phys. {\bf 135} (2009) 217.

\bibitem{ProlhacEM09}
S. Prolhac, M.R. Evans, K. Mallick,  J. Phys. A: Math. Theor. {\bf 42} (2009) 165004.

\bibitem{AritaM13}
C. Arita, K. Mallick, J. Phys. A: Math. Theor. {\bf 46} (2013) 085002.

\bibitem{CrampeMRV15}
N. Crampe, K. Mallick, E. Ragoucy, M. Vanicat, J. Phys. A: Math. Theor. {\bf 48} (2015) 175002.

\bibitem{Vanicat17}
M. Vanicat, J. Stat. Phys. {\bf 166} (2017) 1129. 

\bibitem{FinnRV17}
C. Finn, E. Ragoucy, M. Vanicat arXiv:1712.06809.

\bibitem{Derrida07}
B. Derrida, J. Stat. Mech.: Theor. Exp. (2007) P07023.

\bibitem{BlytheE07}
R.A. Blythe, M.R. Evans, J. Phys. A: Math. Theor. {\bf 40} (2007) R333.

\bibitem{KrebsS97}
K. Krebs, S. Sandow, J. Phys. A: Math. Gen. {\bf 30} (1997) 3165.

\bibitem{Spitzer70}
F. Spitzer, Adv. Math. {\bf 5} (1970) 246.

\bibitem{EvansH05}
M.R. Evans and T. Hanney, J. Phys. A: Math. Gen. {\bf 38} (2005) R195.

\bibitem{Schutz05}
E. Levine, D. Mukamel, G.M. Sch\"utz, J. Stat. Phys. {\bf 120} (2005) 759.

\bibitem{ChatterjeeM17}
A.K. Chatterjee, P.K. Mohanty, J. Phys. A: Math. Theor. {\bf 50} (2017) 495001.

\bibitem{KunibaO17}
A. Kuniba, M. Okado, J. Phys. A: Math. Gen. {\bf 50} (2017) 044001.

\bibitem{KunibaM17}
A. Kuniba, V.V. Mangazeev, Nucl. Phys. B {\bf 922} (2017) 148.

\bibitem{AngelettiTSP13}
F. Angeletti, E. Bertin, P. Abry, IEEE Transactions on Signal Processing {\bf 61} (2013) 5389.

\bibitem{AngelettiEPL13}
F. Angeletti, E. Bertin, P. Abry, EPL {\bf 104} (2013) 50009.

\bibitem{AngelettiJSP14}
F. Angeletti, E. Bertin, P. Abry, J. Stat. Phys. {\bf 157} (2014) 1255.

\bibitem{HarrisRS05}
R.J. Harris, A. R\'akos, G.M. Sch\"utz, J. Stat. Mech. (2005) P08003.

\bibitem{Cappe05}
O. Cappe, E. Moulines, T. Ryden, {\it Inference in Hidden Markov Models},
Springer Ser. Stat. (Springer, New York, 2005).

\bibitem{Evans04a}
M.R. Evans, S.N. Majumdar, R.K.P. Zia,
J. Phys. A: Math. Theor. {\bf 37} (2004) L275.

\bibitem{Evans04b}
R.K.P. Zia, M.R. Evans, S.N. Majumdar, J. Stat. Mech.: Theor. Exp.
(2004) L10001.

\bibitem{Majumdar05}
S.N. Majumdar, M.R. Evans, R.K.P. Zia, Phys. Rev. Lett. {\bf 94} (2005) 180601.


\end{thebibliography}
\end{document}